\documentclass{JHEP3}
\usepackage{amsmath,amsfonts,bbm,euscript}
\usepackage{cite}
\usepackage{epsfig}
\usepackage{graphicx}
\usepackage{psfrag}

\renewcommand{\bar}[1]{
  \overline{#1}
}

\def\unit{1\hspace{-3.6pt}1}
\def\a{\alpha}

\def\t{t}
\def\T{\mathcal T}

\def\st{\star}

\def\w{\wedge}
\def\l{\lambda}
\def\n{\nu}
\def\m{\mu}

\def\ba{\begin{align}}
\def\ea{\begin{align}}
\def\bas{\begin{align*}}
\def\eas{\begin{align*}}
\def\be{\begin{equation}}
\def\ee{\end{equation}}

\def\Tr{\text{Tr}\;}

\def\ap{{\alpha^{\prime}}}
\def\hf{\frac12}
\DeclareMathOperator{\F}{\EuScript{F}}

\def\X{\mathcal{X}}
\def\Y{\mathcal{Y}}
\def\DD{\text{D}\bar{\text{D}}}
\def\DDD{\text{DD}\bar{\text{D}}}

\def\b{\bar}
\def\d{\partial}

\def\DTD{\text{D}{p} + \text{D}(p-2)}

\title{Stability of D1-Strings Inside a D3-Brane}
\author{Louis Leblond, and  S.-H. Henry Tye\\
  Laboratory for Elementary-Particle Physics,
 Cornell University, Ithaca, NY 14853.\\
  E-mail:\email{lleblond@lepp.cornell.edu},
\email{tye@lepp.cornell.edu}}

\abstract{
Within the tachyon condensation approach, we find that a 
D$(p-2)$-brane is stable inside D$p$-branes when the bulk 
is compactified. It is a codimension-2 soliton of the 
D$p$-brane action with coupling to the bulk $(p-1)$-form RR field.
We discuss the properties of such solitons.
They may appear as detectable cosmic strings in our universe.
}

\begin{document}

\section{Introduction}

The evidence that the early universe has gone through an 
inflationary epoch has become very strong. In the brane world 
scenario, where the standard model fields (i.e., photons, 
electrons, quarks, etc., except graviton) are open string 
modes on branes, brane inflation \cite{Dvali:1998pa} is quite 
natural. A particularly simple version of brane inflation involves 
the slow motion of a D3-brane towards an anti-D3-brane 
\cite{Burgess:2001fx,Dvali:2001fw,Alexander:2001ks,Jones:2002cv,
Buchan:2003gx}. 
Recently, based on a realistic superstring compactification where all 
moduli of the vacuum are stabilized \cite{Giddings:2001yu,Kachru:2003aw},
it is shown that this D3-brane pair inflationary scenario may be 
realized in superstring theory 
\cite{Kachru:2003sx,Silverstein:2003hf,Hsu:2003cy,Firouzjahi:2003zy}.
Inflation ends as the D3-brane pair annihilates.
Suppose the standard model fields live in a stack of
(anti-)D3-branes. To allow the energy 
released to heat up the universe to start the hot 
radiation dominated big bang era \cite{Shiu:2002xp,Cline:2002it},
the D3-brane pair annihilation should 
happen close to this stack of branes.

Towards the end of the inflationary epoch, cosmic strings are 
produced 
\cite{Jones:2002cv,Sarangi:2002yt,Jones:2003da,Copeland:2003bj,
Dvali:2003zh,Dvali:2003zj,Binetruy:2004hh,Halyo:2003uu}. 
In the above scenario, the annihilation of the D3-brane 
pair produces cosmic strings that are D1-strings (i.e., D1-branes)
\cite{Sen:1999mg}. 
The tension of such D1-strings are estimated to be 
around $G \mu \simeq 10^{-9}$ to $10^{-10}$ 
\cite{Sarangi:2002yt,Copeland:2003bj}. 
If they survive long enough to evolve as a cosmic string network
\cite{Vilenkin:2000}, they will produce distinct signatures 
that should be detected in the near future, in particular by 
gravitational wave detectors such as LIGO II/Virgo or 
LISA \cite{Damour:2001bk}. 

If the D3-brane collides with a stack of anti-D3-branes and 
annihilates one of them, D1-strings are produced inside 
(or very close to) the (anti-)D3-branes \cite{Sen:1999mg}.
However, in string theory, D1-strings and D3-branes are not BPS 
with respect to each other. In fact, it is generally believed that 
the D1-strings will dissolve inside a D3-brane
\cite{Gava:1997jt,Hashimoto:1997gm}; that is, their energy will 
spread throughout the D3-brane. In this case, the cosmic strings 
would have dissolved almost immediately after they were formed 
and no observable signature will be left.
So the (in)stability of such D1-strings inside
a D3-brane is a very important phenomenological issue. 
Of course, this question in string theory is interesting in its own right. 
 
The possibility of the stability of D1-strings inside
a D3-brane was recently pointed out by Copeland, Myers and Polchinski
(CMP) \cite{Copeland:2003bj}.
The coupling of the RR 2-form field $C_2$ to the Abelian
gauge field $A_1$ inside the D3-brane is finite when the 
extra 6 dimensions are compactifed. This leads to spontaneous symmetry 
breaking and a D1-string inside a D3-brane becomes 
a topologically stable vortex (a D1-vortex) with
localized energy (tension) density inside a D3-brane.
However, this vortex is no longer BPS and has a net zero RR charge 
(but non-zero charge density) as measured by 
$C_2$ inside the D3-brane, since the winding number 
contribution to this charge is canceled by the magnetic flux 
contribution. Because of the conservation of the winding number,
this same vortex becomes a D1-string when moved outside 
a D3-brane, as expected. 

To justify the action used here and in CMP, we present a 
topological argument on the stability of this D1-vortex in the 
context of tachyon condensation, where a 
D1-string also appears as a vortex. Although our approach uses 
boundary superstring field theory, the D1-vortex stability is 
based on topological reasonings and so is insensitive to the details
of the particular framework used.
This approach also reveals the relation between a D1-string 
outside a D3-brane (a BPS vortex due to tachyon condensation) and 
a D1-vortex inside a D3-brane (a vortex due to the $C_2$ coupling). 
Although the dynamics is somewhat involved, this transition as 
a D1-string moves in/out a D3-brane is expected to be smooth.
In the limit of vanishing $C_2$ coupling (e.g., some of
the extra dimensions decompactfy), the size of such a 
D1-vortex grows to infinity. In effect, 
the magnetic flux spreads and a D1-string 
dissolves inside a D3-brane. It is the $C_2$ coupling that 
stabilizes the D1-string inside a D3-brane.

More generally, a D($p-2$)-brane is stable inside 
a stack of D$p$-branes. In a brane world where $(p-3)$ dimensions 
of the D$p$-branes are compactified while the remaining 3 dimensions 
span the universe, D$p$-anti-D$p$-brane inflation would generically
create stable cosmic strings that are D($p-2$)-brane defects (with
$(p-3)$ dimensions compactified).
Their individual stability allows the cosmic string network evolution
and implies that the detection of signatures of cosmic 
strings should be a good test of the brane inflationary scenario;
furthermore, it gives an eagerly sought window to the superstring 
theory itself. Of course, the details of the phenomenology 
may be quite sensitive to the specific inflationary scenario.

The organization of this paper is as follows. In Sec. 2, we review 
the old argument why a D1-brane (i.e., a D1-string) was believed to
dissolve inside a D3-brane. We also review the argument why there 
should be a domain wall enclosed by a D1-string. These two features 
constitute something of a puzzle, as pointed out in CMP. 
We then summarize our resolution to this puzzle. 
In Sec. 3, we solve analytically the vortex solution of 
the model involving the gauge field $A_1$ and $C_2$ inside a D3-brane. 
The $C_2$ coupling to $A_1$ leads to the Green-Schwarz 
mechanism where the gauge field becomes massive. 
We show that this model admits a topologically stable vortex 
solution, which is identified as a D1-vortex, that is, a
D1-string inside a D3-brane, even though this vortex is not BPS. 
We show it has localized energy density, not spread throughout
the D3-brane.
However, the tension of this vortex is logarithmically divergent.  
In Sec. 4, we give the physical picture. 
By comparing to the Abelian Higgs model,
we argue that the divergence in tension is expected in the 
approximate nature of the model, and is easily cured by adding 
the massive mode (namely the radial Higgs field associated to 
the axion) to the low energy effective supergravity action.  
In Sec. 5, we study this problem in the framework of 
tachyon condensation, where we give a topological argument why 
the D1-vortex is stable inside a D3-brane. We also use the boundary 
superstring field theory models to draw the connection between the 
two approaches.

\section{The CMP Puzzle}

In the cosmological context, we want to know if a very large 
D1-string loop can survive over cosmological time scale.
If so, it would interact with other D1-strings and evolve as 
a component in a cosmic string network; it would also oscillate 
and radiate gravitational waves which may be detected. 
On the other hand, if it either dissolves
in superstring/Planck time scale, or shrinks rapidly due to 
the presence of a domain wall, then there 
is no observable signal left for us to detect today.
Let us briefly review the status of a D1-string as known in 
the literature. We explain the CMP puzzle \cite{Copeland:2003bj} 
and then summarize our resolution.

\subsection{Basic Idea of Dissolution}

Consider a D1-string that is either inside a D3-brane, or 
outside a D3-brane. If it is outside, closed string exchange 
between them is attractive, so they will quickly come into
contact. In either case, we are led to study a D1-D3 system. 
It is well known that the $\DTD$ system is a non-BPS state 
with a tachyon in its spectra.  
This state is unstable and there exists
a BPS bound state with the same RR charge but lower mass:
\begin{align}
M_{Dp +D({p-2})} =  \tau_{p}V_{p} + \tau_{p-2}V_{p-2} 
\geq (\tau_{p}^2V_{p}^2 + \tau_{p-2}^2V_{p-2}^2)^{\hf}.
\end{align}
where $\tau_p$ is the tension of a D$p$-brane.
It is argued \cite{Polchinski:1998rq,Polchinski:1998rr,Johnson:2003gi}   
that a D$p$ brane with a constant flux 
throughout its volume has the correct tension and 
charge to be interpreted as the BPS bound state. It is then logical 
to interpret the decay of the $\DTD$ system to its BPS bound state 
as the dissolution of the D$(p-2)$-brane into a constant flux on 
the D$p$-brane, that is, the D$(p-2)$-brane is 'smeared' out 
throughout the D$p$-brane \cite{Gava:1997jt,Hashimoto:1997gm}.
This dissolution means a D$(p-2)$-brane is unstable inside a D$p$-brane.
In the brane world scenario where the D3-branes span our universe,
the D1-string would have dissolved immediately after their production.
If the D1-string intersects with a D3-brane along some 
curve inside the D3-brane, the D1-string is expected to break 
with flux lines connecting the two ends. The spreading of these 
flux lines and the energy density will signal the breaking of 
the D1-string by the presence of the D3-brane.

\subsection{Axionic String and Domain Wall}

The presence of a domain wall bounded by a string was first pointed 
out by Witten \cite{Witten:1985fp}. 
In four large spacetime dimensions,
a D1-string is charged under the 2-form RR field $C_2$, which is 
dual to an axion $\phi$ : $\partial_{\mu} \phi \simeq 
\epsilon_{\mu \nu \rho \sigma} \partial^{\nu} C^{\rho \sigma}$.
That is, $\phi(x)$ increases by $2\pi$ as it circles the D1-string once.
In this sense, D1-strings are axionic strings. 
Since we have not seen a massless axion, we expect the 
Peccei-Quinn $U(1)_{PQ}$ to be broken (say to $Z_k$, $k \ge 1$)
and the axion picks up a mass.
Then the winding of $\phi$ implies a domain wall with
the D1-string as its boundary. To get some idea of the 
domain wall tension, 
assume a potential $M^4(1-\cos\phi)$ that generates an axion mass
$m_{\phi}$ and gives a domain wall tension around
$M^4/m_{\phi}$. Putting in an allowed value for the axion mass 
($m_{\phi} \sim 10^{-14}$ GeV) and assume $M$ to be of the order of 
the superstring scale, we get a very large domain wall tension.
For any reasonable values of $m_{\phi}$ and $M$ (even say 
$M \sim m_{\phi}$), the domain walls confine the D1-strings. 
That is, the D1-string loops will rapidly shrink and disappear, 
leaving no lasting cosmological signatures to be detected.

\subsection{The CMP Puzzle and its Resolution}

As explained above, there are two ways to get rid of the D1-strings as 
cosmic strings in the early universe. However, these two ways are 
incompatible with each other. This is the puzzle pointed out by CMP; 
since the boundary of a boundary is zero, the presence of a domain 
wall implies that a D1-string cannot break. 
On the other hand, if it breaks, i.e., the D1-strings 
develop ends, then the domain wall cannot exist.

Here, we present the resolution to this puzzle. We find that, for 
finite $C_2$ coupling to the Abelian field strength,
the D1-strings do not break. 
In the tachyon condensation approach, where even a D1-string 
outside a D3-brane is treated as a vortex, one simply look at 
the topology/structure of the vacuum degeneracy and argue that 
a D1-vortex is stable both inside and outside a D3-brane.
We see that a D1-string outside a D3-brane is a BPS vortex
due to tachyon condensation while a D1-vortex inside a D3-brane 
is a vortex due to the $C_2$ coupling. However, the actual dynamics 
of the transition between a D1-string and a D1-vortex as it moves 
in/out of a D3-brane is expected to be smooth.

Once its stability is established, one can then write a simple 
effective action to study the properties of such a vortex.
Inside a D3-brane, the axion becomes the would-be Goldstone 
mode that is swallowed by the gauge field inside the D3-brane
via the Higgs (or Green-Schwarz) mechanism. Since
the axion remains massless, the domain wall tension is exactly zero,
i.e., there is no domain wall. At the same time, 
the D1-string becomes a vortex (a D1-vortex) in the Abelian Higgs 
model, with localized energy density (see Fig. 2), 
though it is no longer BPS.
As a consequence, the D1-strings survive inside D3-branes
and they will evolve as a network of cosmic strings in our universe. 
The actual phenomenology of the cosmic string network does depend on 
the details of the inflationary scenario.

As measured by $C_2$, the net RR charge of this D1-vortex 
is zero. Besides a positive contribution to the RR charge
from the winding number of the axion, there is a negative 
contribution to the RR charge coming from the magnetic flux.
This is screening. (However, as shown in Figure 1, these 
two contributions do not cancel locally, 
so there is a non-trivial RR charge density.)
That is, as we move a D1-string inside a D3-brane, it loses its 
RR charge, but retains its winding number.
(One can define this same charge in the Abelian Higgs 
model, and likewise, a vortex there also has a net zero charge.) 
Note that the winding number is identified with the RR charge 
outside the D3-brane. Since it is conserved when a D1-string 
moves inside the D3-brane (it becomes the winding number of 
the D1-vortex), it is a more useful quantum number to keep
track.


\section{The Stability of a D1-String Inside a D3-Brane}

To set up the problem, we first consider a D1-string outside
a D3-brane, i.e., a D1-string as a BPS D1-brane with coupling 
to the 2-form RR field $C_2$.
Next we put it inside a D3-brane. The key new ingredient 
is the coupling of the abelian gauge field $A_1$ to $C_2$.
The Green-Schwarz mechanism takes place and $A_1$ becomes 
massive. This model can be exactly solved where the D1-string 
becomes a vortex in the Abelian Higgs model. This D1-vortex 
remains topologically stable with localized energy density, 
though they are no longer BPS.  There is a very rich
literature \cite{Khare:1995bt}
on Chern-Simons vortices and relation between Chern-Simons terms
and the Abelian Higgs model, though in different contexts. 

\subsection{D1-String Outside a D3-brane}

Let us dimensionally reduce the 10-dimensional theory to an 
effective 4-dimensional theory to get the kinetic term for C$_2$ 
from the bulk action.
Consider $n$ D1-strings coupled to
$C_2$ in the four uncompactified dimensions, all sitting along 
the $z$-axis at $r=|x_{\perp}|=\sqrt{x^2 +y^2}=0$. 
After rescaling $C_2$ to obtain a canonical kinetic term, we have 
(with constant dilaton background):
\begin{align}
S = -\int_{M^4} \hf|dC_2|^2 + 2\pi n a \delta^2(x_\perp)\w C_2.
\end{align}
where $a = \sqrt{2}\tau_1\kappa_4$ 
measures the RR charge of a D1-string whose coordinate
is $\delta^2(x_\perp)dx \w dy$.  $\tau_1$ is the D1-string 
tension and $\kappa_4$ is the 4-dimensional effective gravitational 
coupling. It is related to the 10-dimensional coupling  
$\kappa_4^2 V_6 = \kappa^2=\kappa^2_{10}g_s^2$ where 
$V_6$ is the 6-dimensional compactified volume and 
$g_s$ is the string coupling.  
Introducing the dual of $C_2$, $a d\phi=\st dC_2$, we have
\begin{align}
\label{1out3}
a d\wedge d \phi = d\st dC_2 & = 2\pi n a \delta^2({x}_\perp), \\\nonumber
\st dC_2 & = a d\phi = a\frac{n}{r},
\end{align}
where we have chosen $\phi$ to be dimensionless and $\st = \st_4$ unless
we specify otherwise.
Note that $d \wedge d \phi$ is not identically zero 
since $\phi$ is not single valued;
$\phi(x_{\m})$ increases by $2 \pi n$ as it circles the $z$-axis once.


\subsection{D1-String Inside a D3-brane}

Now consider the same D1-string inside a D3-brane, with
the D3-brane world volume action that involves 
the $C_2$ and the Abelian gauge field $A_1$.
\ba
\label{theaction}
S  = -\int_{M_4} \hf|G_2|^2 + \hf|dC_2|^2+ \xi C_2\w G_2 +  
2\pi na \delta^2({x_\perp}) \w C_2  
\end{align}
where $G_2 = dA_1$, $\xi = \sqrt{2\tau_3}\kappa_4$, $\tau_3$ is 
the D3-brane tension.
Note that $C_2$, $A_1$, $\xi$ and $a$ have dimension of mass.
For $\xi=0$ (e.g., when $V_6 \rightarrow \infty$), 
both $A_1$ and $C_2$ are massless. 
For finite $\xi$, spontaneous symmetry breaking via the 
Green-Schwarz mechanism takes place.

The equations of motion for this action are:
\ba
d\st dA_1 &= \xi dC_2\label{a1eom},\\
d\st dC_2 &= \xi G_2 \label{c2eom} + 2\pi n a\delta^2({x}_\perp).
\end{align}
There are two ways to solve these equations. First let us 
solve for C$_2$. 
The solution of (\ref{c2eom}) is (in non-compact space):
\ba
\label{stdc2}
\st dC_2 = \xi A_1 + a d\phi.
\end{align}
where, by analogy with the solution outside the D3-brane, 
we get the $\delta$ function from the multi-valued $\phi$. 
Putting this back in Eq.(\ref{a1eom}) we get:
\ba
\label{A1eqm}
d\st dA_1 = \st (a \xi d\phi+ \xi^2 A_1).
\end{align}
For the ground state (i.e., $n=0$), where $\phi=0$ (or is a pure 
gauge), we see that $A_1$ has mass $\xi$.
The other way to solve this set of coupled equations
(\ref{a1eom},\ref{c2eom}) is to first solve for $A_1$,
\ba
\label{da1eom}
\st dA_1 = \xi C_2 + d\eta_1 \rightarrow \xi C_2 
\end{align}
since we are considering vortices without 
electric flux, so we expect $d\eta_1=0$.
Putting this back into the equation for C$_2$, we have
\ba
\label{da1eom2}
d\st dC_2 & = (-)\st \xi^2  C_2 + 2\pi n a\delta^2({x}_\perp)
\end{align}
where $C_2$ has mass $\xi$. We can view this system as
\begin{itemize}
\item the gauge field $A_1$ swallowing an axion (the Goldstone boson)
$\phi$ (where $d\phi=\st dC_2 + ...$)
becoming massive: $A_1 \rightarrow A_1 + a d\phi/\xi$;

or

\item  the 2-form field $C_2$ swallowing 
(the dual of) the gauge field $A_1$ 
and becoming massive (where $d\eta_1 = \st dA_1 + ...$):
$C_2 \rightarrow C_2+ d\eta_1/\xi$.
\end{itemize}
Classically, these two dual pictures describe the same
physics, as a massive $C_2$ or a massive $A_1$ has three physical
degrees of freedom in four dimensions.

A comment on the validity of the above effective action 
(\ref{theaction}) is in order.
It is well-known that there is an open string tachyonic mode 
stretching between a D1-string and a D3-brane when they are close
to each other.
It is precisely the presence of such a tachyonic mode that signals 
the instability of the D1-D3 system. This may suggest that, in 
the study of the stability of a vortex solution, we should 
include such a mode in the effective action.
However, it is not clear whether such a tachyonic mode is present 
when we include the induced magnetic flux on the D3-brane, i.e., 
the induced flux may have removed (lifted the mass squared of) the 
tachyon field. To see that there is no tachyonic mode to destabilize 
the solution, we shall turn to tachyon condensation in Sec. 5.  
By including all relevant tachyonic modes in the beginning, 
we examine how spontaneous symmetry breaking take place and keep
track of the tachyon modes and possible solitons. 
The topology of the degeneracy of the final vacuum 
dictates the stability of any possible defect. 
In this approach, a D1-string appears as a BPS vortex and we can 
follow it as it moves inside a D3-brane.
Although tachyon condensation is analyzed in the boundary superstring 
field theory framework, the stability argument is based on 
topology and so should be independent on any of the details. 
There we find that the D1-vortex is topologically stable, justifying 
the above action.  

\subsection{Vortex Solutions Inside a D3-brane}

Here we are interested in any possible vortex solution
of this model. As we shall see, the $A_1$ description (\ref{A1eqm})
is exactly a limiting case of the Abelian Higgs model and the 
physics is completely transparent, while the $C_2$ picture 
(\ref{da1eom2}) is less familiar.
Fortunately, the vortex solution of this model has an explicit
analytic form, so it is relatively easy to clarify the relation
between these two pictures. 
There is a more compelling reason that we prefer to consider
the $A_1$ description (\ref{A1eqm}) as more fundamental. 
This equation allows multi-vortices at different 
locations and orientations as its solitonic solutions. 
In the $C_2$ picture, these vortices are sourced by 
$\delta$-functions input by hand. 

Let us present the solution in details.  
The equation of motion (\ref{A1eqm}) for the gauge field
translates to 
\ba
\label{GmnJ}
\d_\m G^{\m\n} = (a \xi \d^\n \phi + \xi^2A^\n) = J^{\n} 
\end{align}
The solution we are interested in is a vortex with winding number $n$.
Since the winding number is topological and should be independent 
of detailed dynamics, we expect $\phi(x)$ to increase by $2 \pi n$ 
as it circles the $z$-axis once. 
We choose to work in the Coulomb's gauge $\d_\m A^\m= 0$.
So we start with the following ansatz, in cylindrical 
coordinates $(t,z, r, \theta)$ :
\ba
\label{solution}
\phi (x) &= n\theta, \\ \nonumber
A_\theta (x) &= -\frac{an}{\xi r}\a(r)
\end{align}
where $n$ is the winding number of the vortex.
This leads to 
\ba
\label{eomalpha}
\frac{d^2\a}{dr^2} - \frac{1}{r}\frac{d\a}{dr} - \xi^2(\a -1) = 0.
\end{align}
which reduces to the modified Bessel equation.  Let $u=\xi r$,
the solution is just
\ba
\label{asolu}
\a(u) = 1-uK_1(u).
\end{align}
where $K_1(u)$ is the modified Bessel function of order $1$. 
This is valid for all $r \ge 0$. For small $u$:
\ba
K_1(u) & \sim \frac1{u} + \frac{u}{2} \ln(u/2),\\\nonumber
K_0(u) & \sim - \ln (\frac{u}{2})  - \gamma,
\end{align}
and for large $u$,
\ba
K_1(u) \simeq K_0(u) \simeq \sqrt{\frac{\pi}{2u}}e^{-u}.
\end{align}
where we have given the properties of $K_0(u)$ as well.
The following relations are useful (for $u>0$):
\ba
\frac{d}{du}(K_0) = & -K_{1},\\\nonumber
\frac{d}{du}(uK_1) = & -uK_{0}. 
\end{align}
Using Eq.(\ref{solution},\ref{asolu}), we see that
the magnetic field ${\bf B}$ is along the $\hat z$-direction,
\ba
\label{Bfield}
B_z & = \frac{1}{r}\d_r(rA_\theta) 
\\\nonumber
& = -\frac{2 \pi na}{\xi}\delta^2(x_\perp)-an\xi K_0(u)
+\frac{2 \pi na}{\xi}\delta^2(x_\perp) \\\nonumber
     &= -an\xi K_0(u).
\end{align}
where the first $\delta$-function comes from 
the $r^{-1}$ term in $A_\theta$ while the second $\delta$-function 
together with the $K_0$ term comes from the $K_1$ term in $A_\theta$.
It is easy to check this using Stokes' law for each term in 
$A_\theta$.
Integrating over the $xy$-plane, we obtain the total magnetic 
flux of the vortex,
\ba
\int B_z rdrd\theta = - 2 \pi  n \frac{a}{\xi}.
\end{align}

As we shall see, $\xi/a = \sqrt{\tau_3}/\tau_1$ corresponds to 
the electric charge in the Abelian Higgs model. Now we can describe 
this same vortex solution in terms of $C_2$ and $d\eta_1$. 
With the above vortex, only the $(tz)$ component of $C_2$ 
is non-trivial. Using Eq.(\ref{da1eom}) (or integrating 
Eq.(\ref{stdc2})), we have
\ba
\label{c2sol}
C_{01} = -anK_0(u) 
\end{align}
Note that $d\eta_1=0$, as expected, since it is the dual of $B_z$, 
i.e., the electric field, which is absent here.
Alternatively, we can solve for $C_2$ from Eq. (\ref{da1eom2}). 
With only the $(tz)$ component of $C_2$, we have  
\ba 
\partial^2_{\perp}C_{01} =\frac{1}{r}\partial_r(r \partial_rC_{01})
=\xi^2 C_{01} + 2\pi n a\delta^2({x}_\perp)
\end{align}
which is precisely the modified Bessel equation for $K_0$ once we
account for the $\delta$-function correctly  
(similar to the second $\delta$-function in Eq.(\ref{Bfield})). 
So we recover the solution given in Eq.(\ref{c2sol}).
Note that, for large $r$, $C_{01} \sim e^{-\xi r}/\sqrt{r}$.
At short distances, it has a logarithmic singularity 
$C_{01} \sim  -\ln (r)$, that is, it still diverges 
at $r=0$. 
Since $C_2$ falls off exponentially on the codimension 2 plane
inside the D3-brane, the RR charge of the D1-string is screened. This is 
expected since, due to its coupling to $A_1$,  the $C_2$ field is massive;  
the RR charge from $d\phi \simeq 1/r$ is 
canceled by that from $A_1 \simeq -1/r$ (see Fig. 1).  
On the other hand, the winding number is conserved, so we have a consistent
vortex solution of the equation of motion.  As we shall discuss in the
rest of this paper, this vortex solution represents
what a D1-string becomes as it moves inside a D3-brane.
For that reason, we refer to this vortex solution as a D1-vortex. 
\begin{figure}
\begin{center}
\psfrag{a}{$A_\theta(r)$}
\psfrag{b}{$\frac{1}{r}\d_\theta\phi$}
\psfrag{c}{$\st dC_2$}
\includegraphics{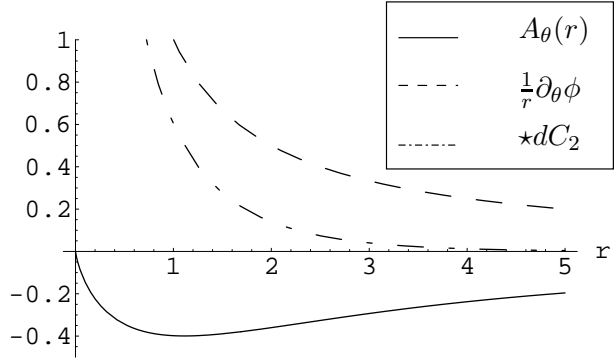}
\caption{$A_{\theta}$ (solid line), $d\phi$ (dashed line) and
$\st dC_2$ (dot-dashed line), where we have set $a=\xi=n=1$.
$d\phi$ is canceled by $A_1$ to screen
the RR charge, so $\st dC_2$ drops exponentially for large $r$.}
\end{center}
\end{figure}


\subsection{The Energy Density of a D1-Vortex}

Let us calculate the tension $\tau$ of the D1-vortex.
We are actually more interested in the energy density in the $xy$-plane.
The localization of the energy density indicates that the D1-string 
has not dissolved completely.
Varying the action (\ref{theaction}) with respect to the $00$ 
component of the metric, we get the energy density of the vortex 
solution,
\ba
\mathcal{E} & = \hf (\partial_rC_{01})^2 + \hf B_z^2 
= \frac{a^2 n^2\xi^2}{2} \left( K^2_1(u)  +  K^2_0(u) \right)
\end{align}
where the topological term $C_2\wedge G_2$ and $C_2 \w \delta^2 (x_{\perp})$
in Eq.(\ref{theaction}) do not contribute since they do not involve the metric.
Integrating over the $xy$-plane, we obtain the tension of the vortex 
\ba
\tau = \lim_{r_s \rightarrow 0}  2 \pi \int_{r_s}^{\infty} 
\mathcal{E} r dr \simeq -\pi a^2 n^2  \ln(\xi r_s) + \frac{2\pi a^2n^2}{4}.
\end{align}
The magnetic field $B_z$ is infinite at $r=0$ but its contribution to 
$\tau$ is square integrable, so we have a finite contribution 
(the second term) to the tension from the magnetic flux (see Fig. 2).
On the other hand, 
the tension $\tau$ diverges in the limit $r_s \rightarrow 0$, due 
to the $(\partial_r C_{01})^2$ term. 
(Note that the vortex from Eq.(\ref{1out3}) also has a divergent tension.)
This divergence is easily interpreted when we compare this system 
to the Abelian Higgs model (see below). 
We need to think of $r_s$ as the inverse of the scalar Higgs mass,
$r_s =1/m_H$. 
In this model, this massive Higgs mode is frozen (or ignored), 
i.e., its mass is effectively set to infinite; this is the origin 
of the above divergence. This divergence is expected to be removed 
when we consider a better effective action.
\begin{figure}
\begin{center}
\psfrag{r}{$r$}
\psfrag{B2}{$B^2$}
\psfrag{E}{$\mathcal{E}$}
\includegraphics{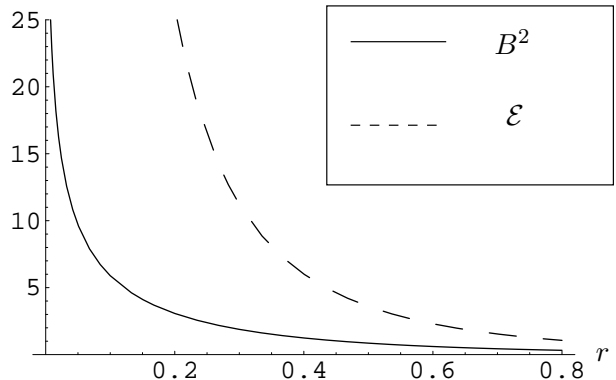}
\caption{The total energy density $\mathcal{E}$ (dashed line)
and the energy density in the B field (solid line). Here, $a=n=\xi=1$.}
\end{center}
\end{figure}

\section{Discussions}

First we show that the above D1-vortex may be viewed as
a special limit of the vortex in the Abelian Higgs model. 
Introducing the analogue of the RR charge, we see that a 
vortex in Abelian Higgs model also has a net zero charge. 
We then summarize the overall physics of the D1-vortex
and comment on how the physics may change as we 
improve the low energy effective theory for this system.
We suggest that the resulting system of $n$ D1-strings 
and a D3-brane is stable as a bound state.
Finally, we show that it is straightforward to generalize 
the stability argument to the case of a D($p-2$)-brane inside
a stack of D$p$-branes. 


\subsection{Relation to the Abelian Higgs Model}

The simplest way (by far) to see the existence and stability of 
D1-vortices in a D3-brane is to realize the direct connection between 
the above vortex solution and a certain limit of the 
Abelian Higgs model. The Abelian Higgs model lagrangian is:
\ba\label{abh}
L = - {\bar {D_\m\Phi}} D^\m\Phi - \frac{1}{4}G_{\m\n}G^{\m\n}
-\frac{\lambda}{4} (\Phi \bar \Phi - v^2)^2
\end{align}
where $D_\m = \d_\m + ieA_\m$.
After spontaneous symmetry breaking, we have a massive gauge 
field and a massive Higgs boson:
\ba
m_A^2=2 e^2v^2,     \quad \quad m_H^2=\lambda v^2.
\end{align}
The equations of motion are
\ba
\label{eqah}
D^\m D_\m \Phi - \frac{\l}{2} (\Phi {\bar \Phi} - v^2) \Phi =  0, \\
\nonumber
\partial_{\mu}G^{\m \n} = 2e Im [{\bar \Phi} D^{\n} \Phi] =  J^{\n}.
\end{align}
Consider a static vortex along the $z$-axis. In Coulomb gauge,
with the ansatz in cylindrical coordinate, 
\ba
A_{\theta} &= -\frac{n}{er} \alpha(r), \\\nonumber
\Phi & = v \beta (r) e^{i \phi},  \quad \quad \phi = n \theta
\end{align}
and the equations of motion reduce to
\ba
\alpha^{\prime \prime} -\frac{\alpha^{\prime}}{r} -
2e^2v^2\beta^2(\alpha-1)=0, \\\nonumber
 \beta^{\prime \prime} +\frac{\beta^{\prime}}{r} -
\frac{n^2 \beta}{r^2}(\alpha-1)^2 -
\frac{\lambda v^2 \beta}{2} (\beta^2-1)=0,
\end{align}
where $\alpha(0)=\beta(0)=0$.
At large $r$, 
\ba
\Phi & = ve^{i n \theta} (1 - c_1 e^{-m_H r} + ...), \\ \nonumber
A_{\theta} &= -\frac{n}{er} (1- c_2 e^{-m_A r} + ...),  
\end{align}
where $c_1$ and $c_2$ are constants that can be determined numerically.
These are the well-known Abrikosov-Nielsen-Oleson vortex solutions 
with winding number $n$ \cite{Abrikosov,Nielsen}.
 
We may introduce the same 2-form field $C_2$ here. 
The current $J^{\n}$ defined in Eq.(\ref{GmnJ}) or Eq.(\ref{eqah})
is conserved: $\partial_{\n}J^{\n}=0$. In terms of $C_2$,
\ba
\epsilon^{\m \n \rho \sigma} \partial_{\n} C_{\rho \sigma}= 
\frac{1}{m_A}J^{\m}
\end{align}
For the vortex solution, we have, for large $r$,
\ba
\epsilon^{\theta r 01}\partial_r C_{01} =
- \frac{i}{m_A} \Phi^{\dagger} \left(\frac{1}{r} \partial_{\theta} \Phi 
+ i e A_{\theta} \Phi\right) \simeq \frac{n}{m_Ar}c_2 e^{-m_Ar}+...
\nonumber
\end{align} 
so 
\ba
\oint \epsilon^{\theta r 01}\partial_r C_{01} rd \theta = 0. \nonumber
\end{align}
That is, in terms of the $C_2$ charge, the 
winding number contribution from $\Phi$ cancels the magnetic flux
contribution from $A_{\theta}$ so that the net $C_2$ 
charge of the vortex is zero, that is, $C_{01}$ drops off
exponentially at large distances from the core. 
For this reason, the $C_2$ charge is not a useful quantum number 
to measure the presence/absence of vortices.
This is identical to the situation in the D1-vortex case. 

Suppose we freeze the Higgs field $|\Phi|=v$ (i.e., $\beta=1$). 
Then we are left with only the axionic field $\phi(x)$ 
(where $\Phi = e^{i \phi}$) and $A_{\m}(x)$, 
and we have only the above equation for $\alpha (r)$,
which reduces to exactly Eq.(\ref{eomalpha}), 
where $\phi$ and $A_{\theta}$ are to be identified, with
$\xi^2 =m_A^2=2 e^2 v^2 $ and $a^2=2v^2$. This allows 
us to identify the D1-vortices as a particular limit of 
the topologically stable Abelian Higgs vortices
(crudely speaking, we may achieve this limit by taking 
$\lambda \rightarrow \infty$). 
The tension of Abelian Higgs vortex is
\ba
\label{Preskill}
\tau =& \int rdrd\theta\left[ D_i\Phi D^i\Phi^\dagger + \hf B^2 + 
V(\Phi)\right], \\\nonumber
 & \simeq~~  2\pi v^2\left[\ln\left(\frac{m_H}{m_A}\right) + 2 + 1 \right],
\end{align}
where the first term represents the gradient energy of the scalar field. 
When we freeze $|\Phi| = v$, we essentially send 
$m_H \sim 1/r_s \rightarrow \infty$,
so the D1-vortex tension diverges logarithmically. This 
logarithmic divergence is what we have obtained earlier.
The presence/variation of $|\Phi|$ smooths out the singularity
and yields a finite tension.
This divergence due to the freezing of $|\Phi|$ is
a consequence of the approximation that can be rectified easily.

There remains a somewhat subtle issue to discuss.  Normally, 
the Abelian Higgs vortex has $\Phi = 0$ at the center of the core,
where the phase (the axion) is not defined and the gauge field is 
massless. In the above model, $A_1$ is massive
everywhere on the brane, so the vortex solution seems to have no `core'.  
Actually, the core should be thought as the point
where the phase is not defined, even with $|\Phi| = 1$ there. 
That is, as long as we remove a point from the $xy$-plane, 
it is no longer simply connected and has non-trivial winding.

\subsection{The Physical Picture}

We have a consistent vortex solution of the D3-brane action with 
the bulk contribution coming from the 2-form RR field. One should
interpret such a vortex as a D1-string inside a D3-brane (see Fig. 2),
as they have the same winding number.
\begin{figure}
\begin{center}
\psfrag{D3}{D3-brane}
\psfrag{vortex}{D1-vortex}
\psfrag{Vortexloop}{D1-vortex loop}
\psfrag{D1}{D1-string}
\includegraphics[height=70mm]{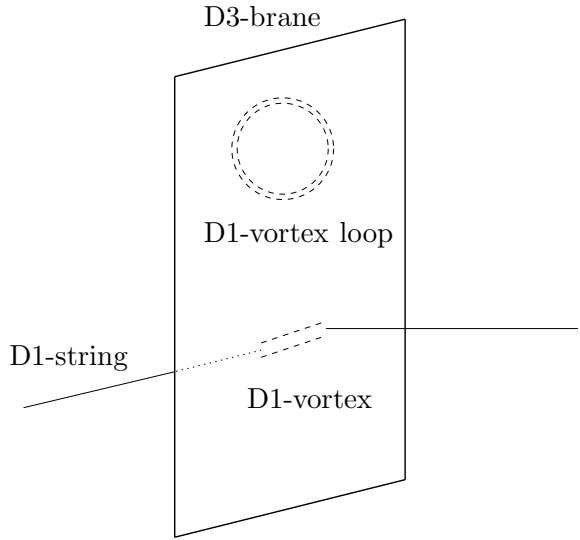}
\caption{A D1-string intersecting a D3-brane and forming a D1-vortex
inside the D3-brane.  The picture also shows a D1-vortex loop 
existing on its own.}
\end{center}
\end{figure}
Outside the D3-brane, both $C_2$ and its dual (the axion $\phi$ 
here) measure the same RR charge of a D1-string, which is identified 
with the winding number. 
Inside a D3-brane, they measure different things. 
$\phi$ again measures the winding number, which is conserved as 
the D1-string moves inside/outside the D3-brane. On the other 
hand, $\st dC_2$ measures the combination of $d\phi$ and $A_1$.
If one uses $C_2$ to measure the RR charge, 
one may view this vortex solution as a combination of two
different vortices : a positively RR charged
vortex ($d\phi \simeq 1/r$) screened by a negative RR
charged vortex ($A_1 \simeq -1/r$), resulting in a net 
zero RR charge as measured by $C_2$. 
This screening of $C_2$ charge happens because $C_2$ becomes 
massive. Naively, they look like
a D1-string screened by an anti-D1-string. However, they cannot
annihilate, since the origin of the RR charge for the
D1-component is different from that for the anti-D1-component.
Furthermore, the two charges do not cancel locally.
As we move this vortex outside a D3-brane, the magnetic flux 
around the core disappears since the gauge field is absent
outside the D3-brane. So this vortex recovers its RR charge
and becomes a D1-string.
In this sense, we believe the conserved winding number is 
the appropriate quantity to follow as we move a D1-string 
inside/outside a D3-brane. 
In the tachyon condensation picture, a BPS 
D1-string is a vortex, so here we simply see that the vortex 
changes its thickness, tension, and magnetic flux as it moves 
in/out a D3-brane,  
but it is always topologically stable, due to the winding 
number measured by the axion $\phi$.

As $\xi \sim g_s^{1/2}$ goes to zero, the vortex grows
(1/$\xi$) to infinite size and its energy density spread
throughout the D3-brane. The D1-string tension ($\xi^{-2}$)
also goes to infinity while the D1-vortex tension grows
only logarithmically. So, as a D1-string moves inside a
D3-brane, it goes to a lower tension vortex with infinite
size, i.e., just like dissolution.

\subsection{Low Energy Effective Theory in String Theory}

The $\delta$-function is the origin of the singular behavior 
of the tension. Including the partner of $C_2$ (i.e., $|\Phi|$)
should get rid of the $\delta$-function singularity.
In string theory, this massive Higgs mode is always present.
Together with $\phi$, they form a complex scalar mode.
Consider the low energy effective theory of a generic Type IIB 
orientifold model. One may write the effective Lagrangian for all 
the string modes as
\be
L=L_{0,2} + L_{0,I}+ L_{M}
\ee 
where $L_{0,2}$ includes only quadratic terms of 
the massless modes, 
while $L_{0,I}$ includes only the interaction terms among the massless 
modes. All other quadratic and higher dimensional operator terms 
are included in $L_{M}$.
The action (\ref{theaction}) is a part of  $L_{0,2}$. 
There is a massive mode $\eta$, with mass $m_{\eta}$, inside $L_{M}$. 
In comparison to the Abelian Higgs model, $\eta$ and $\phi$ form the 
complex field $\Phi=(v+\eta)e^{i\phi}$. 
All terms involving $\eta$ in the abelian Higgs model (and more) 
are included in $L_{M}$. For the ground state, we may ignore $\eta$
or integrate it out. For a vortex solution, since $|\Phi|=0$ at the core,
we must include $\eta$ or $|\Phi|$ in the effective action.
As can be seen in the Abelian Higgs model,
the presence of $\eta$ will smooth out the 
vortex solution and yield a finite tension $\tau$, i.e., the D1-vortex. 
To see this feature more closely within 
superstring theory, one should turn to the tachyon condensation
approach, where D1-strings are treated as vortex defects. 

One may use boundary superstring field theory (BSFT) 
to study the tachyon condensation properties. 
Since the vortex is not BPS inside
the D3-brane, its identification with a D1-string remains to be shown.
The BSFT approach allows us to identify the same vortex outside a 
D3-brane as a D1-string (since it has the correct tension and RR 
charge of a D1-string) and so it is appropriate to consider the 
corresponding vortex inside a D3-brane as a D1-vortex. 

\subsection{Estimate of the D1-Vortex Tension}

We can make a crude estimate of the D1-vortex tension by supposing that 
$r_s \approx \xi^{-1}$. In this case, the energy is almost entirely in the 
$B$ field and we get the tension of the vortex to be:
\begin{align}\label{tension}
\tau_{\text{vortex}} & \simeq \frac{2 \pi a^2n^2}{4} = 
\pi n^2 \tau_1^2\kappa_4^2.
\end{align}
One interesting question is if this solution corresponds to
the BPS bound state of a $n$D1-D3 system.  If not, then we would expect some
kind of instability in the vortex solution.
From SUSY arguments, one can obtain the mass of the BPS bound state for
$n$ D1-branes and a D3-brane:
\begin{align*}
M_{(nD1-D3)} = \sqrt{n^2\tau_1^2V_1^2 + \tau_{3}^2V_{3}^2},
\end{align*}
and the mass of the vortex is therefore expected to be
\begin{align*}
M_{\text{vortex}} = \sqrt{\tau_3^2V_3^2 + n^2\tau_{1}^2V_{1}^2}
- \tau_3V_3 \approx \frac{n^2\tau_{1}^2V_{1}^2}{2\tau_3V_3}.
\end{align*}
So the tension is just
\begin{align*}
\tau_{\text{vortex}} \approx \frac{n^2\tau_{1}^2V_{1}}{2\tau_3V_3}
\approx \frac{n^2\tau_{1}^2}{2\tau_3V_2}.
\end{align*}
Now, what should we take for $V_2$?  
Since the vortex is localized, the correct $V_2$ to take would be the
size of the vortex in the codimension-2 plane, namely 
$V_2 \sim \xi^{-2}$. This yields
\begin{align}
\tau_{\text{vortex}} \approx  \frac{n^2\tau_1^2\xi^2}{2\tau_3}
\approx n^2\tau_1^2\kappa_4^2
\end{align}
which agrees with Eq.(\ref{tension}) up to an $\mathcal{O}(1)$ factors.
Note that the energy of $n$ D1-vortex is proportional to $n^2$, not $n$.
In this sense, a D1-vortex is not BPS. However, the above result 
does suggest that the $n$D1-D3 system saturates the BPS bound and so
is a BPS bound state. This is consistent with the D1-vortex stability 
argument. This means that the solution is not only stable classically
due to its topological nature but also quantum mechanically
(there should be no stringy topological change).

\subsection{Generalization to D($p-2$)-Brane Inside D$p$-Branes}

It is fairly easy to generalize this solution to the D$p$-D($p-2$) system.  
Indeed, 
$\st_{p+1} dC_{p-1} = d\phi$ 
in $p+1$ dimensions and the number of degree of freedom
between $A$ and $C_{p-1}$ match up again.
The action for this system is completely similar to 
what we had before (\ref{theaction}):
\begin{align}
S  = -\int_{M_{p+1}} \hf|G_2|^2 + \hf|dC_{p-1}|^2+ \xi_p C_{p-1}\w G_2 +  
2\pi na \delta^2({x_\perp}) \w C_{p-1}  
\end{align}
where $\xi_p = \sqrt{2\tau_p}\kappa_{p+1}$.

One can generalize this result further to a stack of D$p$-branes.  
In this case, the gauge group is $U(N)$.
Since $C_{p-1}$ couples to the first Chern class,
\begin{align}
\int_{M_{p+1}} C_{p-1}\w \Tr F_2 = \int_{M_{p+1}} C_{p-1}\w G_2
\end{align}
the Green-Schwarz mechanism breaks the $U(1)$ only.
Vortex solutions due to this $U(1)$ breaking are
straightforward extensions of the above solution.

Based on S-duality, we expect similar properties for F-strings 
on a D3-brane. It would be interesting to generalize this result 
to the $(p,q)$-strings. 

To summarize, we have found that, contrary to the usual folklore, a 
D($p-2$)-brane is stable inside a D$p$-brane and it becomes a 
D($p-2$)-vortex. This vortex
\begin{itemize}
\item is topologically stable;
\item is slightly spread, but still has localized energy density;
\item is no longer BPS (when $n$ of them are brought together, their 
tension increase like $n^2$ instead of $n$);
\item becomes a D($p-2$)-brane when moved outside a D$p$-brane;
\item and has net zero RR charge but has a non-zero winding number and 
quantized magnetic flux.
\end{itemize}


\section{D1-String and Tachyon Condensation}

Here we like to study the stability of a D1-string inside 
a D3-brane from the perspective of tachyon condensation, using 
the boundary superstring field theory (BSFT) formalism 
\cite{Kutasov:2000aq,Kraus:2000nj,Takayanagi:2000rz,Jones:2002si}. 
We first show its stability by a topological argument, due to the 
$C_2$ coupling.
Then we relate the tachyon condensation approach to the 
approach discussed earlier.
\begin{figure}
\begin{center}
\psfrag{D1}{D1-string}
\psfrag{TA}{$T_1$}
\psfrag{TB}{$T_2$}
\psfrag{TBINF}{$T_2$ rolls}
\psfrag{TA1}{$T_1$}
\psfrag{Dp2}{D1}
\psfrag{TAROLLS}{$T_1$ rolls}
\psfrag{Vortex}{D1-vortex}
\includegraphics[width=115mm]{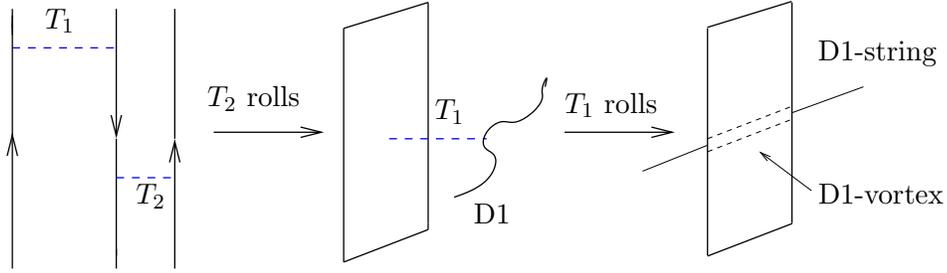}
\caption{The D3-D3-anti-D3 system. The $\DD$-branes on which $T_2$ 
ends are always on top of each other.
Creation of a vortex by first forming the D1-string (from the 
annihilation of the D3-anti-D3 pair) outside the other D3-brane and 
then bringing it back.  
Alternatively, one can form the vortex directly on the D3-brane.
Arrows up represent D3-branes and the arrow down represents the
anti-D3-brane. 
}
\end{center}
\end{figure}
We consider a system of two parallel D3-branes and one anti-D3-brane
(see Figure 4), where one of the D3-branes 
is on top of the anti-D3-brane, with the open string tachyon $T_2$ 
stretching between them. When the other D3-brane is separated from this 
brane-anti-brane pair with separation $\varphi \ne 0$, the annihilation 
of this pair (described by the rolling of $T_2 \rightarrow \infty$) 
produces defects that are D1-strings outside the remaining D3-brane
\cite{Sen:2002nu}. If the separation is small, the remnant of $T_1$
becomes a tachyonic mode stretching between a D1-string and 
the D3-brane. 
When the three branes are all on top of each other ($\varphi=0$), 
any defects produced are inside the remaining D3-brane. 
This model allows us to study the stability of 
a D1-string inside a D3-brane, or whether it can be formed.
This approach also reveals the connection between the D1-string 
and the D1-vortex.

To set up the problem, let us keep track of all scalar modes that
may trigger spontaneous symmetry breaking of the relevant maximum
gauge symmetry $U(2)\times U(1)$.
First, there is the  moduli $\varphi$ that measures the separation between the 
two D3-branes. For $\varphi \ne 0$, $U(2) \rightarrow U(1)_1 \times U(1)_2$.
There is also the tachyon mode $T$ which is a $SU(2)$ doublet and there
is the bulk RR field $C_2$, which couples to a particular 
combination of $U(1)$s. The degeneracy structure of the final vacuum
tells us the existence of any defect.

\subsection{A Topological Argument}

Let us start by considering the setup where all three branes are on 
top of each other (see Fig. 4). The gauge group of this system is 
$U(2)\times U(1) = SU(2)\times U(1)_{DD} \times U(1)_{\bar D}
= SU(2)\times U(1)_Y \times U(1)_L$ 
where the gauge field assignments are (see Table 1 for the various 
definitions of $U(1)$s)  
\begin{align*}
U(1)_{DD}  &\rightarrow B_1, & U(1)_{\bar D} & \rightarrow B_2,
& U(1)_Y & \rightarrow B=(B_1 - \sqrt{2} B_2)/\sqrt{3}.
\end{align*}
The tachyon $T$ couples only to $SU(2)\times U(1)_Y$:
\begin{align}
D_\m T & = \left(\d_\mu + \frac{ig}{2}\left(\begin{array}{cc}
A_3 + B_1- \sqrt{2}B_2 & \sqrt{2} W^+\\
\sqrt{2} W^- & -A_3 + B_1- \sqrt{2} B_2
\end{array}\right) \right)\left(\begin{array}{c}T_1\\T_2\end{array}\right).
\end{align}
The relative normalization (i.e., the $\sqrt{2}$ factor) is fixed by 
moving either of the D3-brane away from the pair 
\cite{Kraus:2000nj,Takayanagi:2000rz}.

First, let us ignore the $C_2$ field.
Without loss of generality, consider the tachyon rolling:
\ba\label{tacprofile}
T = \left(\begin{array}{c}0 \\T_2\end{array}\right).
\end{align}
In BSFT, $T_2 \rightarrow \infty$, though a canonical field 
redefinition brings the $vev$ to a finite value.
So $T_2$ breaks $SU(2)\times U(1)_Y$ to $U(1)_{\gamma}$ with 
gauge field $A$.  
This is just like the spontaneous symmetry breaking in the 
standard electroweak model. So, in the absence of $C_2$, it has 
trivial homotopy and no vortex is possible :
\ba
SU(2)\times U(1)_Y \stackrel{T_2 \neq 0}{\rightarrow} U(1)_\gamma,\\\nonumber
\Pi_1\left(\frac{SU(2)\times U(1)_Y}{U(1)_\gamma}\right) = \unit.
\end{align}
Suppose we naively try to construct a vortex choosing
$T_2 \sim e^{i\theta}$. One may identify this vortex as a 
D1-string inside a D3-brane.  However, this vortex is unstable 
due to the rolling of $T_1$. One may view $T_1$ (actually its remnant)
as the tachyonic D1-D3 open string mode. 
Topologically, the degenerate vacuum is $S^3$, so 
any $S^1$ on $S^3$ will shrink and disappear.

Now, let us introduce $C_2$ and its relevant couplings.
The various field strengths are
\bas
{\mathcal A}^1 & =  \left(\begin{array}{cc}
(A_3 +  B_1)/\sqrt{2} &  W^+\\
 W^- & (-A_3 +  B_1)/\sqrt{2}
\end{array}\right),\\
F^1 &= d{\mathcal A}^1 + {\mathcal A}^1\w {\mathcal A}^1,\\
F^2 &= dB_2,\\
F^- &= F^1-F^2 = d{\mathcal A}^1 + {\mathcal A}^1\w {\mathcal A}^1 - dB_2.
\end{align*}
The relevant coupling terms (between C$_2$ and the gauge field strengths)
in the RR action for the $\DDD$ system are 
\cite{Kennedy:1999nn,Jones:2003ae}
\ba
S_{RR} \propto \int_{M_4} C_2 \w \left(\Tr (F^{-}) + \frac{1}{\t}
(e^{-\lambda\t} -1)
(\Tr(\T F^1 - \t F^2) + ... ) \right)\label{srr}
\end{align}
where $\lambda = 2\pi\ap$ and 
\bas
\t = T^\dagger T = T_1\bar T_1 + T_2\b T_2,\\
\T = TT^\dagger = \left(\begin{array}{cc}
T_1\bar T_1 & T_1\bar T_2\\
T_2\bar T_1 & T_2\bar T_2\end{array}\right).
\end{align*}
For $T = 0$, the above $C_2$ coupling reduces to
\ba
\int_{M_4} C_2\w d(B_1- \sqrt{2}B_2) 
\propto \int_{M_4} C_2\w dB 
\end{align}
So, as discussed earlier, the $U(1)_Y$ is spontaneously broken 
by this coupling, with
\ba
\Pi_1\left(U(1)_Y\right)|_{C_2} = \mathbb{Z}.
\end{align}

\begin{center}
Table 1: $U(1)$s and their gauge fields.

\bigskip

\begin{tabular}{|c|c|}
\hline
$U(1)_{DD}$ & $B_1$\\
\hline
$U(1)_{\bar D}$ & $B_2$\\
\hline
$U(1)_1$ & $A_1 = \frac1{\sqrt{2}}(A^3 + B_1)$ \\
\hline
$U(1)_2$ & $A_2 = \frac1{\sqrt{2}}(-A^3 + B_1)$\\
\hline
$U(1)_Y$ & $B = (B_1-\sqrt{2}B_2)/\sqrt{3}$\\
\hline
$U(1)_\gamma$ & $A = \hf (A^3 + \sqrt{3} B)$\\
\hline
$U(1)^\prime$ & $Z = \hf (-A^3 + \sqrt{3} B)$\\
\hline
$U(1)_C$ & $ \frac1{\sqrt{2}} (A_2 + B_2)$ \\ 
\hline
\end{tabular}
\end{center}

\bigskip

The interesting case is when $T_2$ is non-zero.  
For such a $vev$, tachyon condensation will break the gauge 
group down to $U(1)_\gamma$, as explained earlier.
After tachyon condensation, the $\DD$ pair together with their 
gauge fields disappear. 
As $T_2 \rightarrow \infty$, $Z$ becomes massive and 
decouples, while the other $U(1)_C$ (with gauge field
$(A_2 + B_2)/\sqrt{2}$) of the $\DD$ pair is expected to be 
confined \cite{Bergman:2000xf,Gibbons:2000hf}.
(The confinement of $U(1)_C$ can give rise to electric flux vortices.)
The only gauge field remaining after this tachyon condensation is the 
$U(1)_1$ on the remaining D3-brane.
Putting $T_1 = 0$ but not $T_2$ in Eq.(\ref{srr}) and dropping
the $W^\pm$ fields which play no role here:
\ba
S_{RR} & \propto \int_{M_4} C_2\w(\frac{1}{\sqrt{2}} d(A^3 + B_1)) 
+ C_2\w(e^{-\lambda T_2\bar T_2}d(\frac{1}{\sqrt{2}} (-A^3 + B_1)- 2B_2)) \\\nonumber
& \rightarrow \int_{M_4} C_2\w dA_1 \nonumber
\end{align}
As $T_2 \rightarrow \infty$, $C_2$ couples only to the remaining
$U(1)_1$. So $U(1)_1$ is broken by its coupling to $C_2$.  
The fundamental homotopy group is now 
\ba
\Pi_1\left(U(2)\times U(1)\right) = \mathbb{Z}.
\end{align}
or, more specifically,
\ba
\Pi_1\left(U(1)_1\right)|_{C_2} = \mathbb{Z}.
\end{align}
We conclude that vortex solutions due to the $C_2$ coupling
are topologically allowed and stable, even though the particular
$U(1)$ that is spontaneously broken by $C_2$ changes as the 
tachyon rolls. 
To summarize, of the 3 $U(1)$s, namely 
$U(1)^{\prime} \times U(1)_1 \times U(1)_C$, 
$U(1)^{\prime}$ is Higgsed by $T_2$, 
$U(1)_1$ is Higgsed by $C_2$ and $U(1)_C$ is confined.
There is no destabilizing tachyonic mode left
so a vortex (due to the $C_2$ breaking of 
an Abelian gauge symmetry) is topologically stable as was 
discussed in Section 3. 
This is the so-called D1-vortex, and the 
effective action (\ref{theaction}) is the simplest action that 
captures this key property. 

To obtain a D1-string outside a D3-brane, we turn on $\varphi \ne 0$.
The gauge symmetry is $U(1)_1 \times U(1)_2 \times U(1)_{\bar D}$.
For large enough $\varphi$, $T_1$ is no longer tachyonic (it is a 
normal massive mode). The $D {\bar D}$ pair annihilation yields 
D1-strings as vortices in the bulk, following from the $T_2$-rolling
and
\ba
\Pi_1\left(U(1)^{\prime}\right)|_{T_2} = \mathbb{Z}
\end{align}
where the gauge field of $U(1)^\prime$ is
$Z=(-A^3+\sqrt{3} B)/\sqrt{2}=(A_2- B_2)/\sqrt{2}$.
The identification of these BPS vortices as D1-strings is 
justified as they have the correct tension and RR charge (see below).
As $\varphi$ decreases to around the string scale, $T_1$ becomes 
tachyonic and starts to roll, signaling the instability of a D1-string. 
Magnetic flux on the D3-brane
gathers to screen the RR charge of the D1-string. As 
$\varphi \rightarrow 0$, the D1-string becomes a D1-vortex.
Note that $A_1$ is always present as the key component of the 
Abelian Higgs vortex. 
Although the actual dynamics can be a little complicated, we expect 
the transition from a D1-string (a $U(1)^\prime$ vortex due to 
tachyon condensation) to a D1-vortex (a $U(1)_1$ vortex due to 
$C_2$ coupling) to be smooth.

\subsection{The $\DDD$ System in Boundary Superstring Field Theory}

In principle, one can solve the D1-string to D1-vortex transition using
BSFT. In practice, this is quite complicated. Here, as a modest 
step, let us draw a closer contact between the BSFT approach 
and the effective action (\ref{theaction}) approach.
First, consider the effective action for the D$p$-anti-D$p$-brane pair
in boundary superstring field theory \cite{Jones:2002si}, where
the tachyon is a complex field (not be confused with the tachyon doublet
of the $\DDD$ system).
\begin{align}\label{DDnoGaugeAction}
  S_{(Dp {\bar D}p)} = -\tau_p\int d^{p+1}x\sqrt{-g}\;&2e^{-\lambda T\bar T}
  \F(\X+\sqrt \Y)\F(\X-\sqrt \Y)
\end{align}
where 
\begin{align*}
  \X &\equiv 2\pi\ap^2g^{\mu\nu}\partial_\mu T\partial_\nu\bar T,&
  \Y &\equiv\left(2\pi\ap^2\right)^2
  \Big(g^{\mu\nu}\partial_\mu T\partial_\nu T\Big)
  \Big(g^{\alpha\beta}\partial_\alpha\bar T\partial_\beta\bar T\Big).
\end{align*}
and the function $\F(x)$ is given by\cite{Kutasov:2000aq} 
\begin{align}\label{Fdefinition}
  \F(x) = \frac{4^xx\Gamma(x)^2}{2\Gamma(2x)}
  = \frac{\sqrt{\pi}\Gamma(1+x)}{\Gamma(\hf+x)}.
\end{align}
\begin{align}  
\F(x) = \begin{cases}
    1 + (2\ln2)x + \mathcal O(x^2),&
    0<x\ll1,\\\label{Fseries}
    \sqrt{\pi x}  & x\gg1.
  \end{cases} 
\end{align}
Annihilation of the brane pair happens when $T \rightarrow \infty$.
Tachyon condensation also allows the creation of codimension-2 defects.
For
\ba
T=uz^n=u(x+iy)^n, \quad \quad u \rightarrow \infty
\end{align}
we have a vortex at the origin in the $xy$-plane, whose
tension and RR charge 
are \cite{Kraus:2000nj,Takayanagi:2000rz,Jones:2002si}
\ba
\tau =  4  \pi^2 n \alpha^{\prime} \tau_p
=n \tau_{p-2}, \quad \quad     
\mu_{RR}  = n \mu_{p-2}= n \tau_{p-2} g_s
\end{align}
which are precisely the properties of $n$ BPS D($p-2$)-branes, 
allowing us to identify these defects as D($p-2$)-branes.
These ``vortices'' are BPS with respect to each other, that is,
the total tension of $n$ parallel static vortices 
(with total RR charge $n\mu_{p-2}$) is $n \tau_{p-2}$,
independent of their relative positions.

Next, let us consider the action for the $\DDD$ system.
This action cannot be written in closed form 
because of the mixing terms between the two complex tachyons 
\cite{Jones:2003ae}.
However, for the tachyon doublet profile (large $u$)
\ba
T = \left(\begin{array}{c}0\\T_2\end{array}\right)= 
\left(\begin{array}{c}0\\uz\end{array}\right),
\end{align}
the mixing terms are unimportant and the action reduces to
something very similar to Eq. (\ref{DDnoGaugeAction})
\begin{align}\label{DDDnoGaugeAction}
  S_{(Dp {\bar D}p)} = -\tau_p\int d^{p+1}x\sqrt{-g}\;&(1+
2e^{-\lambda T_2\bar T_2}
  \F(\X+\sqrt \Y)\F(\X-\sqrt \Y)).
\end{align} 
Now, we need to restore the gauge field $A_1$ 
(under which $T$ is neutral)
on the brane. The simplest way would be to add the DBI factor 
such that when $T_2=0$, we get back the DBI action on the 
remaining D3-brane. This gives (keeping only terms to the order we need)
\begin{align}
S_{\DDD} & = -\tau_3\int d^4x \left(\sqrt{-|g+\lambda G_2|} + 
2\sqrt{-g}e^{-\lambda T_2\bar T_2}\F^2\right)
-\frac{1}{4\kappa_4^2}|dC_2|^2\\
& - \mu_3\int_{M_4}\left(C_2\w G_2+ 
i\lambda^2e^{-\lambda T_2\bar T_2}C_2\w dT_2\w d\bar T_2\right)\nonumber
\end{align}
where the arguments of $\F$ are given in Eq.(\ref{DDDnoGaugeAction})
and we have added the relevant terms of the RR action 
\cite{Kennedy:1999nn,Jones:2003ae}.
The last term is the source term for $C_2$. A tachyon profile 
of $T_2=uz$ with $u\rightarrow\infty$ gives us 
a $\delta$-function for this term and this
is the D1-string solution outside the D3-brane. Inside the brane,
a modified $T_2$ may smooth this $\delta$-function.

The equations of motion for this action (expanding the DBI part) are 
\begin{align}
\frac{1}{4\kappa_4^2} d\st dC_2 &= \mu_3G_2 + 
i\lambda^2\mu_3e^{-\lambda T\bar T}dT\w d\bar T,\\\nonumber
\tau_3\lambda^2 d\st dA_1 &= \mu_3 dC_2.
\end{align}
There is also an equation of motion for $T$ but it is somewhat complicated.
Again, if we use the naive profile $T=uz$ with $u\rightarrow\infty$,
we get back the equations of motion in Sect. 3
\begin{align}
i\mu_3\lambda^2 e^{-\lambda T\bar T} dT\w d\bar T & =  
\frac{\mu_1\lambda u^2}{2\pi} 2e^{-\lambda u^2 (x^2+y^2)}dx\w dy\\
& \stackrel{u\rightarrow\infty}{\rightarrow}  \mu_1 \delta^2(x,y)dx\w dy.
\nonumber
\end{align} 
With a different profile for $T$, the solution for both
$A_1$ and $C_2$ will change, but the qualitative feature 
of a vortex is clear. It would be interesting to solve this system
to obtain an explicit tachyon profile.

\section{Remarks and Conclusion}

In this paper we have shown that there exists a consistent
vortex solution of the D3-brane action with RR 2-form bulk terms,
provided the $C_2$ coupling is non-zero. This requires the
compactification of the extra dimensions, as is the case in 
any brane world scenario.
This vortex solution is not BPS, but it has a localized 
energy density, not spread throughout
the D3-brane. As we pull this vortex out of a D3-brane, it
becomes a BPS D1-string; that is, the D1-string is never
broken, although its properties do change as it moves
inside/outside a D3-brane. The transition from a D1-string to a 
D1-vortex may be followed in tachyon condensation.

In terms of tachyon condensation, all D$p$-branes 
(maybe with the exception for $p=9$) may be 
viewed as solitonic defects. 
So there are only odd $p$-defects in 
Type IIB theory. If a D1-string is a boundary of a domain wall, 
or if a D1-string breaks so that it has ends, then such domain 
walls or string ends will appear as even dimensional defects
(2-dim. and 0-dim. respectively), which presumably should not 
exist in Type IIB theory. The resolution proposed in 
this paper is consistent with this belief. 

It will be interesting to study the interaction and intercommutation 
properties of these D1-vortices.
In cosmology, these defects may become cosmic strings in our universe.
We need to determine their properties to find any distinct 
signature that would differentiate them from the usual quantum 
field theory vortices. The importance
of this study cannot be overstated since this is, at the 
moment, a most promising window into superstring theory and 
the inflationary scenario.



\section{Acknowledgment}

We thank Sarah Buchan, Hassan Firouzjahi, Nick Jones, Saswat Sarangi, 
Benjamin Shlaer and especially Joe Polchinski 
for many useful discussions. This material is based
upon work supported by the National Science Foundation under Grant
No.~PHY-0098631.

\providecommand{\href}[2]{#2}\begingroup\raggedright\endgroup

\bibliographystyle{JHEP}

\end{document}